\documentclass[12pt,a4paper]{article}
\usepackage[dvips]{graphicx}
\newcommand{\bea}{\begin{equation}}
\newcommand{\eea}{\end{equation}}
\newcommand{\ber}{\begin{eqnarray}}
\newcommand{\eer}{\end{eqnarray}}

\newcommand{\p}{\partial}
\newcommand{\da}{\delta}

\newcommand{\f}{\frac}

\renewcommand{\r}{\right}
\newcommand{\lef}{\left}

\newcommand{\epsa}{\epsilon}

\begin{document}
\title{Harmonic oscillator states as oscillating localized structures near Hopf-Turing instability boundary}
\author{A. Bhattacharyay\footnote {Email: arijit@pd.infn.it} \\ Dipartimento di Fisika 'G. Galilei'\\ Universit\'a di Padova\\ Via Marzolo 8, 35131 Padova\\Italy\\}

\date{\today}
\maketitle
\begin{abstract}
A set of coupled complex Ginzburg-landau type amplitude equations which operates near a Hopf-Turing instability boundary is analytically investigated to show localized oscillatory patterns. The spatial structure of those patterns are the same as quantum mechanical harmonic oscillator stationary states and can have even or odd symmetry depending on the order of the state. It has been seen that the underlying Turing state plays a major role in the selection of the order of such solutions.\\  
PACS number(s): 87.10.+e, 47.70.Fw
\end{abstract}
\par
Localized spatio-temporal structures play a very important role in the long time dynamics of many chemical and hydrodynamic systems. Investigation of local instabilities is an important field [1-8], specially the so called topological defects in chemical and biological systems [9-15] which are believed to play a role in the transition from phase to defect chaos regime. Interaction of local and global modes give rise to very complicated large scale dynamics [16-19]. It is still an unsolved problem as to how dissipation and nonlinear feedback balance each other to produce localized oscillations (oscillons) and in that context some field theoretic studies [20-24] are worth mentioning. So, it makes a very rich and broad field - the interaction of very small and large scales- in the context of dynamical systems. In an extended system, the formation of spatio-temporal defects, which are generally localized in nature on a stationary or uniformly moving frame, are generally believed to be generated as a result of local and global symmetry mismatch which can arise from boundary conditions, multi-stability of nonlinear states etc. On the contrary, intrinsic local instabilities of a system with its own symmetry can act as a symmetry breaking agent of the global state its embedded in and are very important in this respect. By intrinsic we mean those which have not been generated by multi-stability of global states such as a droplet of one state pinned by suitable front structures within an extended one. Rather, these are local instabilities to the global states. Such instabilities can grow up locally and be connected to a global state by fronts or pulses whose symmetry reflects that of the localized core. Thus, its about thinking the other way where basic symmetry of the localized core that determines the symmetry of the large scale structures.    
\par
In the present work we are going to show a class of new local instabilities which are like harmonic oscillator states and can arise in oscillatory as well as steady form. The most interesting thing about this class of local structures is their spatial symmetry which can be even and odd depending on the order of the state. Since these local structures are Hermite polynomial in their spatial profile, they can be symmetric or antisymmetric depending upon the order of the polynomial. In the present work it has been shown that instabilities of order greater than zero can grow depending upon the situation prevails. Such symmetric and antisymmetric local structures are also shown to be a steady state of the system with relatively larger spatial extent. We will show that for such a local instability to grow, a coupling of the Turing and the Hopf modes are essential when the bifurcation is supercritical. A pure Hopf environment can also produce oscillatory as well as steady (more restricted) form of such localized structures. Although, in the following we mainly work in a region of phase space close to a Hopf-Turing instability boundary but the result can be generally applied to any situation governed by a complex Ginzburg-Landau equation having the real part of the coefficient of the diffusive term to be negative. In a purely Turing region, it will be shown that, the instability will appear in a non-oscillatory form and a coupling to Hopf mode is essential for it to oscillate.

\par
The slowly varying amplitudes of Turing and Hopf instabilities being denoted by $T$ and $H$ the coupled amplitude equations in one dimensions are given by 
\ber\nonumber
\f{\p T}{\p t} &=& \epsa a_t T + b_t\f{{\p}^2 T}{{\p x}^2} -c_t{\vert{T}\vert}^2 T -d_t{\vert{H}\vert}^2T \\
\f{\p H}{\p t} &=& \epsa a_h H + b_h\f{{\p}^2 H}{{\p x}^2} -c_h{\vert{H}\vert}^2 H -d_h{\vert{T}\vert}^2H \\\nonumber
\eer
Where all the coefficients with a suffix h are generally complex and those with suffix t are real. This system has a steady state of the form $T = T_0 e^{ikx} $ and $ H = H_0 e^{i\omega t}$ where the wave number and the oscillation frequency are given by
$$
k^2=\f{\epsa a_t -c_t{\vert{T_0}\vert}^2 -d_t{\vert{H_0}\vert}^2}{b_t}
$$and,
$$
\omega = \epsa a_{hi} -c_{hi}{\vert{H_0}\vert}^2- d_{hi}{\vert{T_0}\vert}^2
$$ where the suffix $hi$ represents the imaginary part of the complex coefficients.
 Let us see what happens to local perturbations which has a Gaussian profile to these steady states. To see that, we put the perturbations of the form $\da T e^{-x^2/2b}$ and $\da H e^{-x^2/2b}$ on the Turing and Hopf states respectively to get
\ber\nonumber
 \f{\p \da T}{\p t} &=& K_1 \da T + b_t\lef [\f{{\p}^2 \da T}{{\p x}^2} -\f{2x}{b}\f{\p \da T}{\p x}\r ] -d_tH_0T_0Cos(\omega t)e^{ikx}\da H \\
\f{\p \da H}{\p t} &=& K_2 \da H + b_h\lef[\f{{\p}^2\da H}{{\p x}^2}-\f{2x}{b}\f{\p \da H}{\p x}\r ] -d_hH_0T_0Cos(kx)e^{i\omega t}\da T \\\nonumber
\eer
where
\ber
K_1 &=& \epsa a_t -\f{b_t}{b} + b_t\f{x^2}{b^2} - c_t{T_0}^2 -d_t{H_0}^2\\
K_2 &=& \epsa a_h -\f{b_h}{b} + b_h\f{x^2}{b^2} - c_h{H_0}^2 -d_h{T_0}^2\\\nonumber
\eer
The region of investigation is such that $x^2/b^2$ is very small and we will neglect it. Let us re-scale the space co-ordinate such that $x = \f{x}{\sqrt{b}}$ and thus we are looking at a somewhat bigger scale with $b>1$ and then make some reasonable approximations. The second approximation is - region of local structures are so small compared to the long wavelength of the Turing order $k$ we will replace $Cos(kx)$ and $e^{ikx}$ by unity. This approximation is justified because comparatively large scale modes are supported by amplitude equations. With these simplifying assumptions and rescaling the linearized model looks like
\ber\nonumber
 \f{\p \da T}{\p t} &=& K_1 \da T + \f{b_t}{b}\lef [\f{{\p}^2 \da T}{{\p x}^2} -2x\f{\p \da T}{\p x}\r ] -d_tH_0T_0Cos(\omega t)\da H \\
\f{\p \da H}{\p t} &=& K_2 \da H + \f{b_h}{b}\lef[\f{{\p}^2\da H}{{\p x}^2}-2x\f{\p \da H}{\p x}\r ] -d_hH_0T_0e^{i\omega t}\da T \\\nonumber
\eer 
where now $K_1$ and $K_2$ are devoid of terms containing $x^2/b^2$.
To separate space and time let us take the solution in the form 
\ber\nonumber
\da T &=& T(t)H_n(x)\\
\da H &=& H(t)H_n(x)\\\nonumber
\eer
where $H_n(x)$ is the Hermite polynomial of $nth$ order. Putting this in the above linearized Eq.5 and dividing them by $\da T$ and $\da H$ respectively we get
\ber\nonumber
 \f{1}{T(t)}\f{\p T}{\p t} &=& K_1  + \f{b_t}{bH_n(x)}\lef [\f{{\p}^2 H_n(x)}{{\p x}^2} -2x\f{\p H_n(x)}{\p x}\r ] -\f{d_tH_0T_0}{T(t)}Cos(\omega t)H(t) \\
\f{1}{H(t)}\f{\p H}{\p t} &=& K_2  + \f{b_h}{bH_n(x)}\lef[\f{{\p}^2H_n(x)}{{\p x}^2}-2x\f{\p H_n(x)}{\p x}\r ] -\f{d_hH_0T_0}{H(t)}e^{i(\omega) t}T(t) \\\nonumber
\eer 
Each of the above two equations now can be separated in the space and time parts. After we separate the space and time part of above equations, we see that the space parts are clearly decoupled and the coupling exists only in the time part. the decoupling of the space part is very important in view of the fact that we can now set them equal to some arbitrary constants and let them be $-2n\f{b_t}{b}$ and $-2n\f{b_h}{b}$ respectively for the Turing and Hopf case where $n$ is an integer. 
The space parts would now look like
\ber\nonumber
\f{b_t}{bH_n(x)}\lef [\f{{\p}^2 H_n(x)}{{\p x}^2} -2x\f{\p H_n(x)}{\p x}\r ] &=& -\f{b_t}{b}2n\\
\f{b_{h}}{bH_n(x)}\lef[\f{{\p}^2H_n(x)}{{\p x}^2}-2x\f{\p H_n(x)}{\p x}\r ] &=& -\f{b_{h}}{b}2n\\\nonumber
\eer
 Thus, Eq.8 makes it visibly clear that in each part the equation for Hermite polynomial solution is satisfied and there is a spatial instability in the form of a harmonic oscillator state $e^{-x^2/2}H_n(x)$ (for $b=1$) and the condition for its growth is to be determined.

The temporal parts give coupled linear equation as
\ber\nonumber
\f{\p T(t)}{\p t} &=& \lef(K_1-\f{2nb_t}{b}\r )T(t) -d_tT_0H_0Cos(\omega t)H(t)\\
\f{\p H(t)}{\p t} &=& \lef(K_2-\f{2nb_{h}}{b}\r )H(t) -d_{h}T_0H_0T(t)e^{i\omega t}\\\nonumber
\eer
Now we would like to see whether the spatial solutions of the local structures in the form of various bound states of a quantum mechanical harmonic oscillator grows or not. We know that $K_1 = b_t k^2-\f{b_t}{b}$ and $K_2 = -\f{b_{hr}}{b}+i\omega$ where the second relation comes from the steady state condition of the Hopf mode. With $K_1$ and $K_2$ replaced by appropriate expressions Eq.9 looks like
\ber\nonumber
\pmatrix{\dot{T(t)}\cr \dot{H(t)}\cr }=\pmatrix{b_t(k^2-\f{2n+1}{b}) & -d_tT_0H_0Cos(\omega t) \cr -d_{h}T_0H_0e^{i\omega t} & -b_{h}\f{2n+1}{b}+i\omega \cr
}\pmatrix{{T(t)}\cr {H(t)}\cr }\\
\eer

 Thus, with the coupling absent a steady instability will grow in the amplitude of the Turing mode under the condition
\ber
k^2 \ge \f{2n+1}{b}
\eer
where the wave number of the preexisting rolls decides on the upper limit of the order of localized structure given the fixed spread of it. Although we see here that an instability of lowest order $n=0$ has maximum linear growth, but that may not be the only selection criterion for the order of the instability. Other factors and nonlinearity might play in selecting a nonzero order when they come in. When $b_t$ is negative the situation reverses and the wave number $k$ gives here the lower bound to the order $n$ and we can safely say that in the case of growth of an instability in the Turing part when
\ber
k^2 \le \f{2n+1}{b}
\eer
 In such a situation the nonlinearity must play a role in putting the upper bound to the order of the instability since linear growth is proportional to $n$. Since the bifurcation is sub critical with negative $b_t$ which is more effective in producing nonzero order of the instability, the transition can show hysteresis. Now, what the coupling of Turing and Hopf mode does is making the steady instability developed in pure Turing situation oscillate and obviously producing some sort of frequency entrainment with the underlying Hopf mode following the standard theory of forced linear systems.
\par
In the same decoupled situation with the Hopf part of the equations we see that it also admits growth of such an oscillatory instability separately but only when the real part of $b_h$ is negative. With the real part of $b_h$ equal to zero the growth of a steady instability is marginal with an order $n$ satisfying the relation $\f{b\omega}{b_h}=2n+1$. This is a rather stringent condition which allows a discreet set of $\omega$ of the preexisting Hopf mode to produce steady local structures. Thus, a continuous band of $\omega$ is effective in making grow an oscillatory instability with an oscillation frequency 
\ber 
\bar{\omega}=(2n+1)\f{b_{hi}}{b}-\omega 
\eer 
and it would be more commonly observed in situations governed by a single one dimensional complex Ginzburg-Landau equation. No coupling with the Turing mode is necessary here to produce such an oscillatory instability to grow but on the other hand the coupling to hopf mode is essential for an instability to oscillate which originates from the underlying Turing state. Also in the Hopf part the instability of higher order has a higher growth rate and one would expect to see some with nonzero $n$ and the other factors have to play a role on the selection of the upper bound to the order. 
         
\par
The important fact about these instabilities is that they are intrinsic in nature and can appear separately in Turing or Hopf environment. At the same time, a mixed Hopf-Turing phase is also locally unstable to such instabilities. In a large $b$ limit such steady harmonic oscillator state like local structures are a solution of the system when one could neglect the nonlinear terms away from the origin where $e^{-x^2/b}$ is very small and thus, a region away from the origin but not too far will be appropriate for the appearance of these steady structures. In such a situation the order of the steady state is given by $\epsa a_t = (2n+1)\f{b_t}{b}$ and similar relation will come from the Hopf part by separately equating real and imaginary coefficients. Thus, the Hopf part is capable of producing bi stability of such steady local structures when the real and imaginary parts differ in magnitude. However, a detailed treatment of such symmetric and antisymmetric local instabilities giving rise to moving pulse or front structures or surviving in the presence of inhomogeneous oscillatory global environment can prove to be very interesting due to the clash of symmetries. Its the spatial symmetry of these structures which can be odd and would definitely make the coexistence of them in a spatially symmetric global state an interesting thing to study. When oscillates, such antisymmetric localized structures can also influence the dynamics of its surroundings in a nontrivial way and further work in the field of interaction of these localized structures with its environment can reveal such effects.   

\end{document}